\let\latexaddtocontents\addtocontents
\documentclass[a4paper,twocolumn,11pt, accepted=2024-03-07]{quantumarticle}
\let\addtocontents\latexaddtocontents
\pdfoutput=1
\usepackage[utf8]{inputenc}
\usepackage[english]{babel}
\usepackage[T1]{fontenc}
\usepackage{amsmath}
\usepackage{hyperref}
\usepackage{amsfonts}
\usepackage{mathtools}
\usepackage{physics}
\usepackage{dsfont}
\usepackage{verbatim}
\usepackage{color}
\usepackage{tikz}
\usepackage{lipsum}

\usepackage{textgreek}
\usepackage{amssymb}
\usepackage{eucal}
\usepackage{caption}
\usepackage[normalem]{ulem}
\usepackage{wasysym}
\usepackage[backend=bibtex, sorting=none]{biblatex}
\begin{document}

\title{Dissipation as a resource for Quantum Reservoir Computing}

\author{Antonio Sannia}
\affiliation{%
 Institute for Cross-Disciplinary Physics and Complex Systems (IFISC) UIB-CSIC, Campus Universitat Illes Balears, 07122, Palma de Mallorca, Spain.
}%

\author{Rodrigo Mart\'inez-Pe\~na}
\affiliation{%
 Institute for Cross-Disciplinary Physics and Complex Systems (IFISC) UIB-CSIC, Campus Universitat Illes Balears, 07122, Palma de Mallorca, Spain.
}%

\author{Miguel C. Soriano}
\affiliation{%
 Institute for Cross-Disciplinary Physics and Complex Systems (IFISC) UIB-CSIC, Campus Universitat Illes Balears, 07122, Palma de Mallorca, Spain.
}%
\author{Gian Luca Giorgi}
\affiliation{%
 Institute for Cross-Disciplinary Physics and Complex Systems (IFISC) UIB-CSIC, Campus Universitat Illes Balears, 07122, Palma de Mallorca, Spain.
}%

\author{Roberta Zambrini}%
\affiliation{%
 Institute for Cross-Disciplinary Physics and Complex Systems (IFISC) UIB-CSIC, Campus Universitat Illes Balears, 07122, Palma de Mallorca, Spain.
}%

\maketitle

\begin{abstract}
 Dissipation induced by interactions with an external environment typically hinders the performance of quantum computation, but in some cases can prove to be a useful resource. In the field of quantum reservoir computing, we show the advantages induced by dissipation when tunable local losses are introduced into models of spin networks. Our approach, based on continuous dissipation, is compared with existing quantum reservoir computing models based on discontinuous erasing maps. A clear improvement in the computational capabilities of the system is shown, as tested in different benchmark tasks involving linear and nonlinear memory as well as forecasting capacity. The effect of finite ensembles is also addressed. Finally, we formally prove that, under non-restrictive conditions, our dissipative models form a universal class for reservoir computing. This means that they can approximate any fading memory map with arbitrary precision.
\end{abstract}

\section{Introduction}
Quantum science holds promise to revolutionize the technological perspectives in fields such as computing \cite{NAP25196,PRXQuantum.1.020101}, communication \cite{Gisin2007}, sensing \cite{RevModPhys.89.035002}, or cryptography \cite{cryp}.
In particular, the introduction of quantum logic allows the development of algorithms that outperform the corresponding classical ones \cite{Harrow2017,Shor,grover,deutsch1992rapid, bernstein1997quantum,
}, with broad and interdisciplinary applications in chemistry \cite{QChem}, finance \cite{QFinance,stamatopoulos2020option} and machine learning \cite{biamonte2017}. Still, the experimental
implementation of these algorithms is a challenging task with state-of-the-art technology \cite{Preskill2018quantumcomputingin,NISQ}. Currently, available devices mostly rely on a few (up to hundreds) of noisy qubits, an obstacle that must be overcome to achieve a quantum advantage in ``real world'' computational problems or quantum error correction. 

A common problem is that dissipation, caused by the interaction between qubits and the external environment, produces decoherence phenomena that hinder fragile quantum resources. Dissipation, however, can be turned into a positive computational resource in some cases as for dissipation engineering,  which exploits the system-environment interaction as an integral part of the computation process \cite{Eng_diss} and for quantum memories \cite{Eng_diss2}. 
Applications range from quantum control \cite{Koch_2016} to non-equilibrium quantum thermodynamics \cite{vinjanampathy2016quantum,Manzano}, quantum biology \cite{huelga2013vibrations}
or quantum synchronization \cite{manzano2013synchronization,cabot2018unveiling}. The aim of this work is to establish the beneficial effect of tunable dissipation on quantum machine learning and, in particular, on the field of quantum reservoir computing (QRC) \cite{Mujal2021}. 

Reservoir computing (RC) is a machine learning method rooted in Recurrent Neural Networks and is particularly suited for time series processing \cite{lukovsevivcius2012reservoir}. Classical RC was initially proposed as Liquid State Machines \cite{LSM} and Echo State Networks \cite{ESN} and later generalized to include a broad spectrum of physical implementations \cite{PRC}.  Two important advantages of RC are the simplicity of the training process, which requires only linear regression optimization rather than the gradient descent procedures of traditional neural networks, and the ability to multitask. RC is now well-established as a powerful analog neuromorphic method in classical machine learning. Successful experimental realizations of temporal processing with RC and applications have been reported in the last years, including electronic, spintronic, and photonic systems 
\cite{BookRC,moon2019temporal,grollier2020neuromorphic,van2017advances}.

 Recently, this supervised machine learning approach has been proposed in quantum substrates \cite{Mujal2021} to solve both classical \cite{Nakajima,Spin1,Chen_2019,Spin2,martinez2020information, Ridb, kalfus2022hilbert, nokkala2021gaussian, Boson4, Qcircuit, QCirc2,NakaDiss, QPhoton} 
 and quantum tasks \cite{Qtask1, Fermion,Fermion1, Qtask3, Qtask4,Qtask5, Nokkala2023}, in platforms as diverse as spin networks \cite{Nakajima,Spin1,Chen_2019,Spin2,martinez2020information, Ridb}, fermionic setups \cite{Fermion,Fermion1}, continuous-variable bosonic oscillators \cite{kalfus2022hilbert, nokkala2021gaussian, Boson4,Nokkala2023}, superconducting qubits quantum computers \cite{Qcircuit, QCirc2,NakaDiss} and photonic integrated circuits \cite{QPhoton}.
The main motivation for this burgeoning interest is that the superposition principle allows quantum reservoirs to reach an exponential advantage over the classical ones in terms of the number of degrees of freedom. 
Evidence of this feature has been observed and quantified in QRC through the Information Processing Capacity \cite{dambre2012information} with spins \cite{martinez2020information} and with Gaussian networks as well \cite{nokkala2021gaussian}. However, taking advantage of all the degrees of freedom in a real experiment, as discussed in Refs. \cite{mujal2022time,garcia2022scalable,Hu2023} and in Sec. \ref{Subsec:measurments}, is a non-trivial task due to the limitations imposed by the finite number of measurement samples available.

In spite of the variety of dynamical systems that can serve RC purposes, there are some key features needed for temporal series processing in this architecture (formally defined in Sec. \ref{sec:univ}). In a nutshell, the system is required to store some past input information (fading memory) and to forget its initial conditions (echo state property). For quantum RC, it is known that the presence of dissipation is a necessary feature \cite{Nakajima,Chen_2019} and, consequently, all the proposed QRC frameworks contain some dissipation.

The pioneering work of Fuji and Nakajima (FN) is based on a spin-network QRC scheme \cite{Nakajima} and a discrete erase-and-write map. This alternates a unitary evolution followed by an instantaneous input encoding (resetting one qubit). The erasure can be realized via a local dissipation restricted to the input node(s) and modeled by a Lindblad master equation (Sec. \ref{sec:QR} and Appendix \ref{App:Approximation}). An alternative paradigm for QRC that we propose is based on the presence of engineered losses for each of the spin nodes, giving rise to continuous dissipation (CD), acting at all times, and on a continuous input drive obtained by tuning an external magnetic field. We will show that the degree of adaptability of such a tailored dissipation allows the system to optimize its performance according to the task faced, thus outperforming the FN model, where the control over dissipation is very low. Furthermore, we will demonstrate that the CD model achieves universal QRC, which means that any generic task to be solved with RC can be arbitrarily well approximated by considering only these kinds of models. 

The paper is structured as follows: in Sec. \ref{sec:QR}, we review the FN model and introduce the CD one, discussing its universality in Sec. \ref{sec:univ}; in Sec. \ref{sec:memory}, we analyze the memory properties of such models while in Sec. \ref{sec:forecast} we test them in time-series forecasting tasks; in Sec. \ref{sec:experiments}, we study the effects of limiting the number of measurement samples on the performance of the systems, and we present possible experimental platforms for implementing the CD model; finally, discussion and conclusions are given in Sec. \ref{sec:conclusions}.

\section{Quantum reservoirs}\label{sec:QR}
According to the RC theory, a general model of a reservoir must satisfy some necessary conditions to operate. One is the echo state property, which consists of the disappearance of the dependence of the initial condition on the reservoir dynamics over time \cite{EC_prop}. This is a necessary feature because otherwise, the training phase would also have to consider the initial state choice, making the whole procedure inefficient. Furthermore, a proper RC system needs fading memory to process the information of a time series without requiring an infinite amount of physical resources \cite{fading_mem}. Therefore, in the quantum case, dissipation is a crucial feature to satisfy these two conditions \cite{Chen_2019}. Finally, an RC model must also be able to discriminate between different input sequences in order to adapt its behavior to the particular problem of interest \cite{separation}. We will formalize and contextualize all these features for our case study in Appendix \ref{sec:universality}.
In this section, we will consider two different quantum reservoir computers focusing on the role played by dissipation in their functioning and performance. In all the cases, we will consider spin network-like reservoir systems, whose output layer is given by a linear combination of local and/or global observables. Moreover, as we will show, all the models presented are designed to process discrete-time signals.

Let us start our discussion by recalling the first model of QRC introduced by Fuji and Nakajima (FN) in Ref.~\cite{Nakajima} and based on a transverse-field Ising model characterized by the Hamiltonian 
\begin{equation} \label{eq:Ham_Naka} 
    H=\sum_{i<j}^{N}J_{ij} \sigma_{i}^{x}\sigma_{j}^{x}+h\sum_{i=1}^{N}\sigma_{i}^{z},
\end{equation}
where the $i,j$ label the sites of the network, $\sigma_{i}^{a}$ $(a=x,y,z)$ are the Pauli matrices acting on the $i$-th site, $h$ is the value of the homogeneous magnetic field and $J_{ij}$ is the spin-spin coupling following a uniform distribution in a pre-determined interval $[-J_s, J_s]$. 
We will consider a real input sequence of length $M$, $\{s_k\}_{k=1}^{M}$, and rescaled, so that  $s_k \in [0,1]$ $\forall k$. The FN updating rule of the reservoir is obtained by feeding the input to the state of one qubit of the network, for the sake of the definiteness we say the first one. In particular, the state of the first qubit is prepared in an input-dependent coherent superposition
$\rho^{(1)}_{k}=\ket{\psi_{s_k}}\bra{\psi_{s_k}}$ where $\ket{\psi_{s_k}}=\sqrt{1-s_k}\ket{0}+\sqrt{s_k}\ket{1}$ (see \cite{mujalJPC} for a discussion of this and different encoding effects).
Then, the system unitarily evolves for a certain interval of time $\Delta t$, only following the dynamics generated by the Hamiltonian $H$. The complete update rule is:
\begin{equation}\label{eq:Naka}
    \rho_{k+1}=e^{-iH\Delta t} \rho_{k+1}^{(1)}\otimes\Tr_{(1)}\{\rho_{k}\} e^{iH\Delta t}
\end{equation}
where $\Tr_{(1)}\{\cdot\}$ is the partial trace over the first qubit and
$e^{-iH\Delta t}$ is the time evolution operator, assumed as unitary between the input injections. Still, the map \eqref{eq:Naka} exhibits dissipation and decoherence and this occurs instantaneously at each input injection (as modeled by the partial trace). 

A natural way to experimentally realize such injection on a NISQ device consists of realizing a measurement on the first spin and, subsequently, setting its state with a quantum gate conditioned by its outcome. 
In digital implementations like the IBM quantum computer, the reset of qubits after a measure is a recently implemented feature \cite{Qiskit} although it is rather slow and then susceptible to uncontrollable decoherence \cite{deco}.   

Let us instead consider an alternative QRC approach characterized by interactions with an external environment under Markovian conditions. The most general time evolution of a density matrix is described by the Gorini-Kossakowski-Sudarshan-Lindblad (GKLS) Master Equation \cite{breuer2002theory,lindblad1976generators, gorini1976completely}:
\begin{equation} \label{eq:Linb}
\dot{\rho}=\mathcal{L}[\rho] \equiv -i[H,\rho]+\sum_{i}\gamma_{i} ( L_{i}\rho L_{i}^{\dagger}-\frac{1}{2} \{L_{i}^{\dagger}L_{i},\rho \})
\end{equation}
where $\{\gamma_{i}\}$ are the decay rates of the qubits in each external environment and the operators $\{L_{i}\}$, called jump operators, identify the environment action on the qubits.
In Eq. \eqref{eq:Linb}, we identify unitary $\mathcal{U}$ and dissipative $\mathcal{D}$ superoperators:
\begin{equation*}
\dot{\rho}=(\mathcal{U}+\mathcal{D})[\rho]
\end{equation*}
with $\mathcal{U}[\rho] =-i[H,\rho]$ and $\mathcal{D}[\rho]$ equal to the remaining (Lindbladian) term.
We will consider dissipation leading to local losses (i.e. independent losses at each of the $N$ reservoir nodes) modeled by N jump operators $L_{i}=\sigma_{i}^{-} \equiv \frac{1}{2}(\sigma_{i}^{x}-i\sigma_{i}^{y})$. We notice that this model adequately accounts for Markovian dissipation in independent baths whenever the interaction between reservoir nodes is weak (for a more accurate discussion see \cite{Cattaneo_2019}). Furthermore, protocols to engineer a Lindbladian with these characteristics are also known \cite{Eng_diss}. In Appendix \ref{App:Approximation}, we show how dissipation acting locally on the first oscillator leads to an evolution that is able to reproduce the FN map (\ref{eq:Naka}).
 
\begin{figure}
    \includegraphics[width= \linewidth, keepaspectratio]{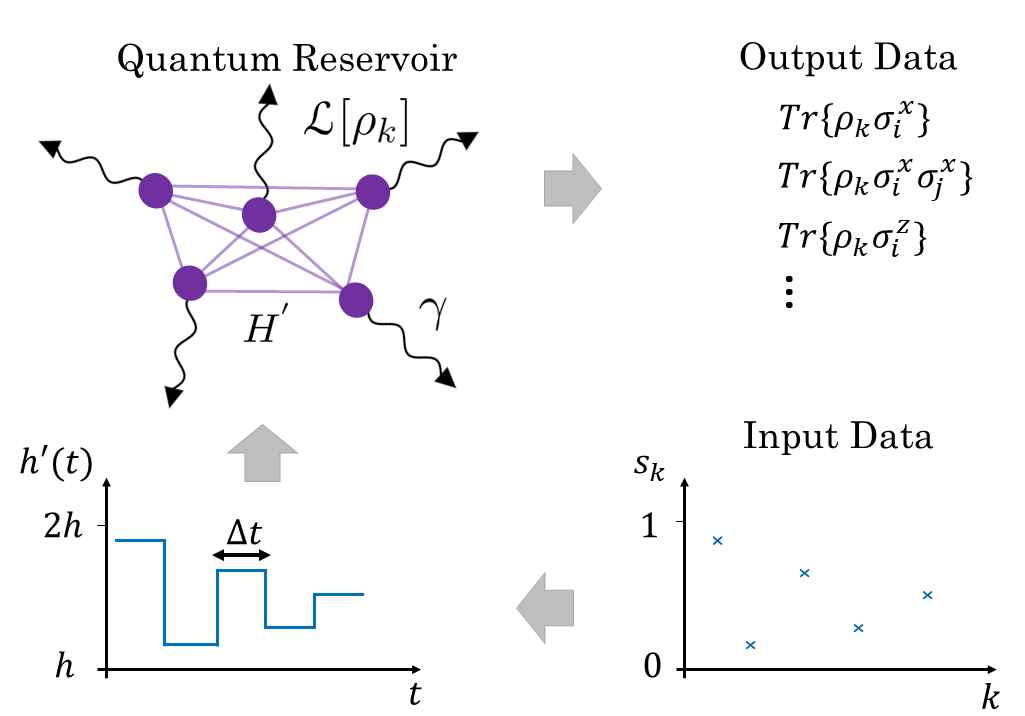}
    \centering
    \caption{Schematic of the CD model proposed in Eq. \eqref{eq:New_model}: spin network (with Hamiltonian $H'$ given in Eq. \eqref{eq:Our_Ham}) with uniform losses at rate $\gamma$ governed by $\mathcal{D}_L$. The discrete input (bottom right) is injected uniformly into the network nodes through a time-dependent magnetic field (bottom left). The output layer is constructed by measuring a portion of the observables of the density matrix of the reservoir.}
    \label{fig:OM}
\end{figure}

Let us now introduce a different model of quantum reservoir computing characterized by continuous dissipation (CD) where the input is injected into the system through temporal driving, varying a Hamiltonian parameter. 
We consider, in particular, a variable magnetic field modulated in the x-direction into the Hamiltonian \eqref{eq:Ham_Naka}:
\begin{equation} \label{eq:Our_Ham} 
    H^{'}=\sum_{i<j}^N J_{ij} \sigma_{i}^{x}\sigma_{j}^{x}+h\sum_{i=1}^{N}\sigma_{i}^{z}+h'(t)\sum_{i=1}^{N}\sigma_{i}^{x}, 
\end{equation}
where the spin-spin coupling coefficients are randomly chosen from a uniform distribution in $[-J_s,J_s]$ as in Eq. (\ref{eq:Ham_Naka}). The time-dependent $h'(t)$ encodes the input time series and coherently modifies the evolution of the reservoir. For each input $s_{k} \in [0,1]$, driving persists for a certain interval $\Delta t$ according to the assignment rule 
$h'_k(t)=h\cdot(s_k+1)$. As dissipation is required for QRC \cite{Chen_2019}, the unitary dynamics generated by the Hamiltonian (\ref{eq:Our_Ham}) will not be sufficient for our purposes.
The simplest kind of dissipation we can introduce consists of adding local, uniform losses to the reservoir nodes ($\gamma_{i}=\gamma$ $\forall i$).  We will see that the decay rate $\gamma$ strongly affects the memory and computational properties of the system. During each time interval, the reservoir dynamics is governed by
\begin{equation}
  \dot{\rho}= -i[H_k^\prime,\rho]+{\cal D}_{L}[\rho] 
  \label{Eq:ME_CD}
\end{equation}
where the local (L) dissipator is given by ${\cal D}_{L}[\rho]\equiv\gamma \sum_{i} ( \sigma_i^{-}\rho  \sigma_i^{+}-\frac{1}{2} \{ \sigma_i^{+}\sigma_i^{-},\rho \})$ and where $H_k^\prime=H^\prime(h'_k)$. 
Therefore, in this case, $\mathcal{U}$ and $\mathcal{D}_{L}$ are functions of the input and $\gamma$ respectively and the updating rule of the reservoir is:
\begin{equation}
  \label{eq:New_model}
    \rho_{k+1}=e^{[\mathcal{U}(s_{k+1})+\mathcal{D}_L(\gamma)]\Delta t} \rho_{k}\equiv e^{\bar{\mathcal{L}}(s_{k+1})\Delta t}\rho_{k}
\end{equation}
where the specific choice of continuous-dissipation and input-driven dynamics is distinguished by the bar in $\bar{\mathcal{L}}$.
A schematic representation of this CD model is shown in Fig. \ref{fig:OM}.

\section{Universality}
\label{sec:univ}

As anticipated in Sec. \ref{sec:QR}, a reservoir computing model, to properly work, must fulfill a set of necessary features, namely the echo state property \cite{EC_prop}, fading memory \cite{fading_mem} and input separability \cite{separation}. These properties clearly determine the class of problems that this machine-learning paradigm is designed to solve. 
A reservoir computing model is universal if it exhibits these properties. Then, in principle, its outputs can reproduce any fading memory map arbitrarily well \cite{LSM,grigoryeva2018echo,Qcircuit,nokkala2021gaussian}. Indeed, as in other contexts, universality refers to the capability of a class of systems (reservoir computers here) to
approximate any map in a much larger class, with arbitrary
precision.

As one of our main results, in Appendix \ref{sec:universality}, we prove that the proposed CD model has the universality property. Previous results of universal QRC in other settings have been reported in Refs. \cite{Chen_2019,nokkala2021gaussian}.   {Beyond the specific model proposed here, } our proof can be applied to a wider class of quantum reservoirs that evolve according to a master equation. In short, we showed that if a generic input-dependent generator $\mathcal{L}(s_{k})$ (i) admits only one stationary state, (ii) is a continuous function of $s_k$ and (iii) takes different stationary states for different inputs, then the corresponding quantum reservoir model, for a value of $\Delta t$ sufficiently long, is universal. Interestingly, these general conditions can be used for finding new quantum reservoir models even beyond spin network implementations.

\section{Computational benchmark tasks\label{sec:tasks}}

We now proceed to numerically evaluate the computing performance of the FN and CD models. We will tackle two families of benchmark tasks that usually appear in the RC literature, namely memory and forecasting tasks.
We set the reservoir to $N=5$ spins, which implies a density matrix with a number of elements $\sim 4^5$. This reservoir size allows us to achieve reasonably good performance in all considered tasks, while still not requiring excessive numerical resources and being accessible on a standard desktop computer. The effect of increasing or decreasing the reservoir size is discussed in Appendix \ref{Sec:Scaling}.

For the readout layer, we choose a linear combination of the expectation values of Pauli strings with length  one ($\langle \sigma_{i}^{a}\rangle$)  and  two
($\langle \sigma_{i}^{a}\sigma_{j}^a \rangle $), with  $a=x,y,z$, and $ 1\le i,j \le N$, $i \neq j$. Altogether, the output layer is made of $45$ nodes which are the parameters determined during the training phase. This set of observables is adequate for solving all the tasks we have tackled with high accuracy while utilizing a large number of degrees of freedom that scales polynomially with the number of qubits.   

In all the cases studied, we have subjected the systems to a washout phase before carrying out a training and an evaluation phase. It consists in letting the reservoir evolve for a certain number of time steps in order to guarantee the ESP \cite{mujal2022time}. We have found that a number of $1000$ time steps is always a sufficient value for the washout. The subsequent $1000$ points of the dynamics, 
were used to train the free weights of the output layer (training phase) \cite{BookRC}, while the next $1000$ points were used to evaluate the performance (evaluation or test phase). We have numerically verified that the data sets employed in all the tasks considered are sufficiently long to avoid overfitting and to collect meaningful statistics to properly evaluate performance. More details about how we trained the reservoirs of interest can be found in Appendix \ref{Sec:training}.

\subsection{Memory tasks}\label{sec:memory}
In this section, we will present the performance of tasks related to the capacity of the systems to linearly and nonlinearly process the memory of previous inputs, starting with a linear memory test: the short-term memory task (STM) \cite{STM}. Following the standard procedure, the input originates from a uniform random distribution in the interval $[0,1]$ and the expected target at each time step ($y_k$) is to reproduce the previous input for a given delay $\tau$
\begin{equation*}
    y_{k}=s_{k-\tau}.
\end{equation*}
Indeed, the STM task measures the ability of a system to store information about the input received a certain number $\tau$ of time steps in the past, which is an indicator of linear memory.
The metric chosen to evaluate all the presented memory tasks is the capacity:
\begin{equation*}
    C=\frac{\mathrm{cov}^{2}(\textbf{y},\bar{\textbf{y}})}{\sigma(\textbf{y})^{2}\sigma(\bar{\textbf{y}})^{2}}
\end{equation*}
where $\textbf{y}$ and $\bar{\textbf{y}}$ are respectively the time series of the targets and the predictions, $\mathrm{cov}(\cdot)$ is the covariance and $\sigma(\cdot)$ is the standard deviation. The coefficient $C$ ranges between $C=0$ (complete mismatch of the predictions) and $C=1$ (perfect accuracy). 

Our analysis encompasses a coarse-grained exploration to optimize the hyperparameters of the models for all the tasks discussed in this paper. Working in the units of $J_s$, the degrees of freedom $h$, $\Delta t$, and $\gamma$ are varied by orders of magnitude, taking values from the following set: $\{0.01, 0.1, 1, 10 \}$. For each possible combination of these hyperparameters, we simulated 100 different random pairs of coupling sets $\{J_{ij}\}$ and input sequences and took the metric average over them as a representative value. Finally, the combination with the average of the maximum performance in the given task was considered optimal.

\begin{figure}[!t]
    \includegraphics[width= \linewidth, keepaspectratio]{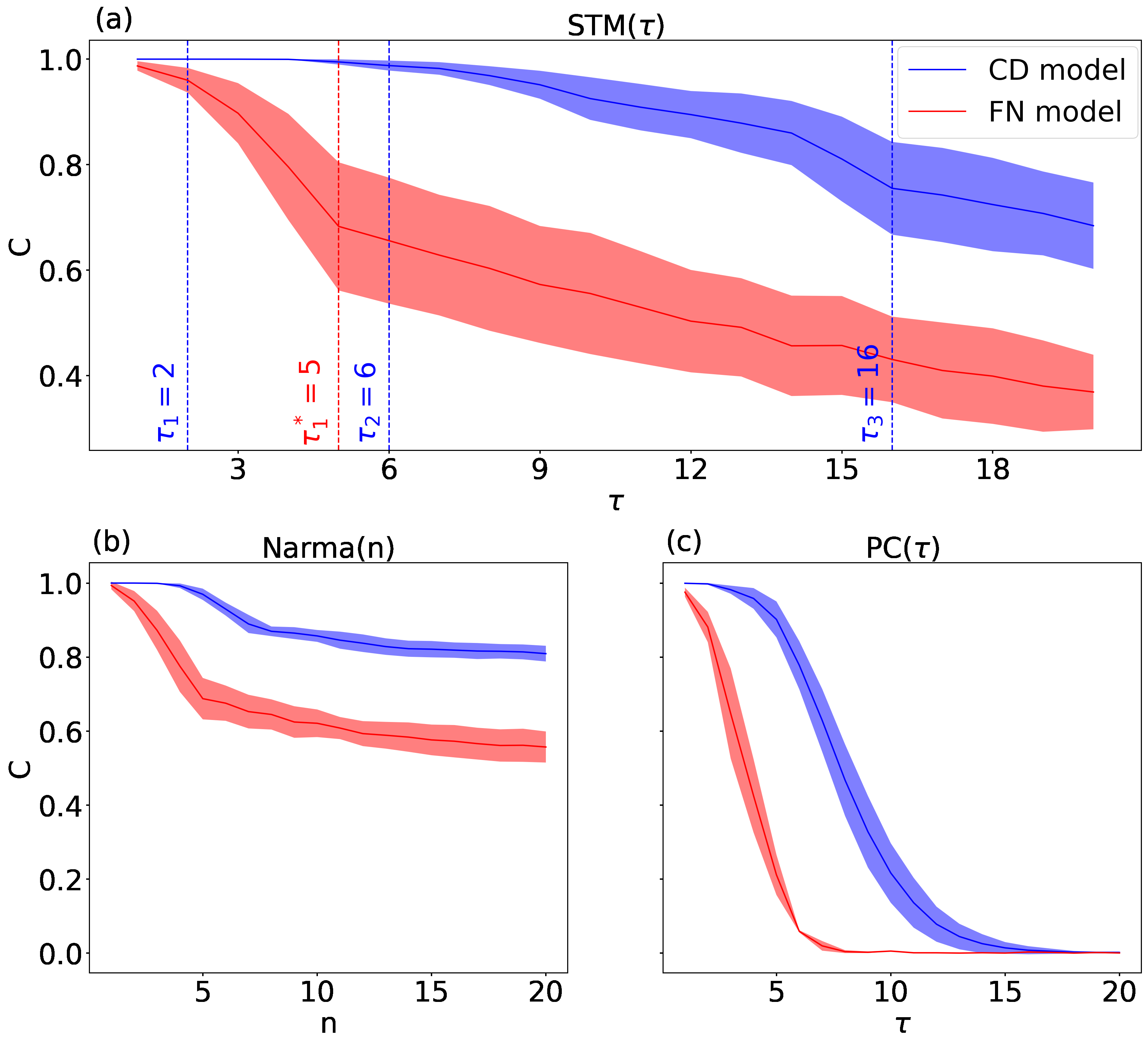}
    \caption{Performance evaluation of different memory tasks
    for the FN model described in Eq. \eqref{eq:Naka} (red) and for CD in Eq. \eqref{eq:New_model} (blue). Capacity for  
  STM task (a),  NARMA task (b), and parity check (c).
  Optimized hyperparameters as specified in the main text for the STM task, while for NARMA  and parity check the hyperparameters are also optimized but not reported, being not relevant for the conclusions. A number of 1000 time steps have always been used for the washout, training, and evaluation phases. In all the plots the shadow regions cover one standard deviation taken as statistical error over the 100 random realizations of spin coupling values for the Hamiltonians of the systems $\{J_{ij} \}$. }
    \label{fig:Mem}
\end{figure}

The results of the STM task, specifying the values of the hyperparameters, are shown in Fig. \ref{fig:Mem} (a). In general, for different orders of delay, the optimal set of free parameters varies for all the models, which makes it evident that the learning capability of the reservoir is strongly influenced by the choice of such hyperparameters.  We found that FN reservoir memory is maximized for $h = 0.1$ and $\Delta t = 10$ when $\tau < \tau_1^{*} = 5$ while in the complementary case $\Delta t$ and $h$ change to 1. For our proposal (CD model  \eqref{eq:New_model}) four different optimal sets have been found. Indicating the free hyperparameters through a triple of the form $(h, \Delta t, \gamma)$, for the regions of delay $\tau < \tau_1 = 2$, $\tau_1 \le \tau < \tau_2 = 6$, $\tau_2 \le \tau < \tau_3 = 16$ and $\tau_3 \le \tau \le 20$ the optimal sets are respectively (0.1, 10, 1), (0.01, 10, 0.1), (0.01, 10, 0.01) and (0.01, 1, 0.1).
For all the values of the delay, the CD dynamical map \eqref{eq:New_model} is able to reach better performances than the FN model. This suggests that the tunable damping rate introduced in the dynamics, which is an additional free hyperparameter with no equivalent in the FN map, allows us to have more control over the linear dependence from the past injected input. 

The generality of this result is also addressed in Appendix \ref{Sec:Scaling} where we report the STM achieved at delay 10  for a smaller ($N=3,4$) and a larger ($N=6,7$) number of qubits. Our results indicate that the advantage in the performance of the CD model is sustained.

Another well-established and challenging benchmark test studied in the context of RC is the nonlinear auto-regressive moving average (NARMA) task, first introduced in the context of RNNs \cite{Narma}. This task, beyond requiring linear memory, also adds the requirement of non-linear memory, and its formulation depends on a given order of delay $n$. At any time step, the NARMA$(n)$ target is
\begin{eqnarray}
    y_{k}&=&  0.3y_{k-1}+0.05y_{k-1}\sum_{j=1}^{n}y_{k-j}  \nonumber\\
    &+&1.5s_{k-n}s_{k-1}+0.1.
\end{eqnarray}
As usual, to prevent divergences until an order of delay equal to $20$, it is necessary to define the target of the task rescaling the input in the interval $[0,0.02]$. The NARMA task performances, for both systems, are shown in Fig. \ref{fig:Mem} (b). Also in this case the results show that the CD model performs better than the FN one, broadening our conclusions about the STM task, also for a target that includes some nonlinearity.

We conclude our analysis of the memory properties considering the parity-check task \cite{PC}. In this case, we work with a binary random input sequence $s_k \in \{0,1\}$. The desired output 
for a delay $\tau$ is given by
\begin{equation*}
    y_{k}=\sum_{j=1}^{\tau}s_{k-j} \mod 2.
\end{equation*}
This task is strictly non-linear and, in fact, for a delay equal to $\tau$, it has the degree of nonlinearity of a monomial of degree $\tau$ of the previous inputs.
In order to reach performances not negligible for $\tau > 1$, we have introduced in both the QRC models the time multiplexing, considering $V=15$ virtual nodes. It means that, for each input injection, during the evolution time $\Delta t$, the observables in the readout have been collected not only in the final time but in $V$ equidistant times of the dynamics \cite{Nakajima,martinez2020information}. The optimal performances for this task are shown in Fig. \ref{fig:Mem} (c). Even for this fully nonlinear task, the CD model outperforms the FN one.

Therefore, from the results obtained, we can conclude that the CD model, with the introduction of the new tunable dissipation $\gamma$, reaches better performances for all the memory tasks considered. The proposed approach of continuous dissipation allows indeed to achieve better control of the dynamical system response. We also remark that the optimal results for the FN model are in agreement with the ones theorized in \cite{martinez2021dynamical}, where the role of the QRC performance in connection with dynamical phase transitions was discussed. However, for the CD case, the optimal hyperparameters cannot be related to the same physical phenomena. In fact, the CD model does not present any dynamical phases and, as expected, we have found different optimal values of $h$ and $\Delta t$ for the tasks considered. In this case, the physical interpretation is related to the interplay between the unitary and dissipative parts of the system evolution.

\subsection{Time series forecasting}\label{sec:forecast}

Apart from memory tasks, RC is especially suited for chaotic time series forecasting. A popular benchmark application
is the prediction of the well-known Mackey-Glass (MG) dynamical evolution \cite{MackeyGlass}. The target obeys  the following differential equation:
\begin{equation}
 \label{eq:MG}
    \dot{s}(t) = -0.1s(t)+\frac{0.2s(t-\tau)}{1+s(t-\tau)^{10}}
\end{equation}
being in the chaotic regime for $\tau=17$, as reported in previous works on time series forecasting \cite{farmer1987predicting,jaeger2004harnessing}. 

During the training phase, we have injected, as input sequence into the system, the numerical solutions of Eq. \eqref{eq:MG} sampled with a time resolution $t_r=3$ (see \cite{Mg_2} for more details) and with input values rescaled in the interval $[0,1]$. The readout weights were trained to solve the one-step-ahead prediction task:
\begin{equation*}
    y_k=s_{k+1}
\end{equation*}
During the evaluation phase, the systems have been left to evolve in an autonomous way taking, at each time step, their previous prediction as a current input. 

\begin{figure}[!t]
    \includegraphics[width= \linewidth, keepaspectratio]{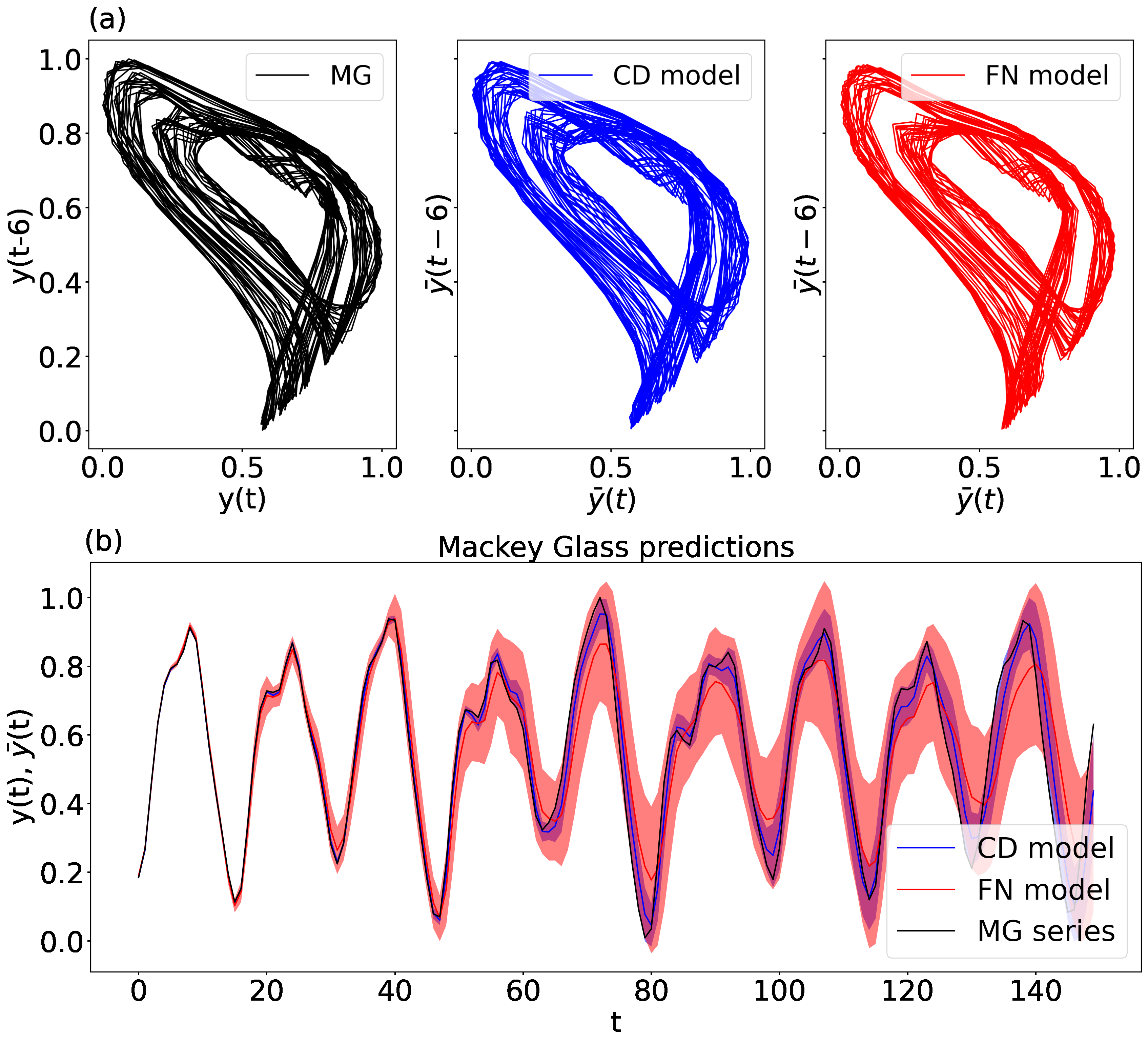}
    \caption{Results of the Mackey-Glass time-series prediction of the systems. In all the cases we have set 1000 points for the washout and training phase. The optimal free hyperparameters found, and fixed in the following plots, are $h = 1$, $\Delta t =10$ for FN model and $h = 0.1$, $\Delta t =0.1$, $\gamma = 10$ for the CD model. (a) Chaotic attractor autonomous reproductions (CD and FN) in the phase space $y(t)$-$y(t-6)$ compared to the target attractor (MG) for one exemplary realization of the systems couplings $\{J_{ij} \}$. 2000 points have been used for the plots. (b) Predictions of 150 points in the autonomous phase contrasted to target values of the series averaged over 100 random realizations of the coupling values. The shadows cover one standard deviation taken as a statistical error. 
                                                            }
    \label{fig:MG}
\end{figure}

Here, we can distinguish between the short-term (weather-like) forecast of the dynamical trajectories and the long-term (climate-like) forecast of the chaotic attractor \cite{pathak2017using}.
We first show in Fig. \ref{fig:MG} (a) that both the CD and FN models are able to reproduce the shape of the chaotic attractor of the Mackey-Glass time series (climate-like forecast).
In order to perform a quantitative analysis, we have also tested the capabilities of the systems to predict the oscillations of the target trajectory from the start of the autonomous phase (weather-like forecast). The metrics used to evaluate the performances for this task was the mean squared error:
\begin{equation*}
    MSE =  \frac{\sum_{i=1}^{M}(\bar{y}_i-y_i)^2}{M}
\end{equation*}
where the index $i$ iterates the times and $M$ is the number of points analyzed. Setting $M=150$, we were able to see differences as plotted in Fig. \ref{fig:MG} (b).  
The average MSE values for the CD and FN models over these 150 samples are $ 8 \cdot 10^{-3}$ and $5 \cdot 10^{-2}$, respectively.
This last result allows us to conclude that also for the forecasting task explored here, the introduction of local losses makes it possible to achieve higher performance.

\section{Experimental feasibility}\label{sec:experiments}

In this section, we discuss relevant aspects of the experimental implementation of the CD model on real hardware. First, we consider the case where access to a number of measurement samples is limited, going beyond ideal conditions, \cite{mujal2022time,garcia2022scalable,Hu2023}, and address the achievable performance. Then, we will present some potential experimental platforms where the proposed QRC model can be implemented.

\subsection{Measurement effects}\label{Subsec:measurments}
\begin{figure}[t]
    \includegraphics[width= \linewidth, keepaspectratio]{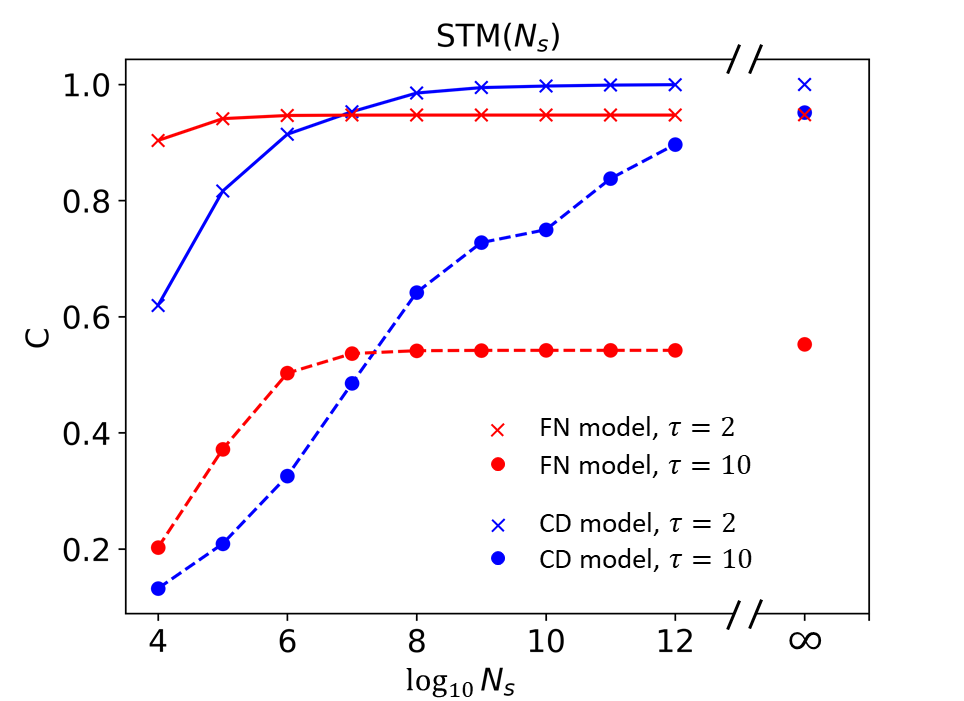}
    \caption{STM task performance as a function of the number of samples $N_S$. Two delay values are considered, $\tau = 2$ and $\tau = 10$, while varying the magnitude of $N_S$ from $10^4$ to $10^{12}$. The right part of the plot shows the ideal performance.}
    \label{fig:Ns}
\end{figure}
 Access to the information contained in the dynamics of a quantum system is generally limited by the stochastic nature of quantum mechanics. In the benchmark tasks presented so far, our analysis has been performed in the ideal case where the readout layer is built up from the exact expectation values of selected observables. In a real experiment, however, only finite measurement samples can be accessed. We will now show how the performance of the two QRC models under consideration (FN and CD) is affected by this fundamental limitation. In particular, denoting by $N_s$ the number of samples (i.e. the size of the available ensemble), we will construct the output layers of the models by approximating each observable of interest $\langle \hat{\mathcal{O}} \rangle$ with the mean of its measurement:
\begin{equation*}
    \langle \hat{\mathcal{O}} \rangle_{N_s} = \frac{\sum_{i = 1}^{N_s} \mathcal{O}_i}{N_s}
\end{equation*} 
where we use $\mathcal{O}_i$ to indicate a generic value of the measurement. 
We follow the same approach as in \cite{mujal2022time}: considering that $N_s >> 1$,  we can apply the central limit theorem so that each $\langle \hat{\mathcal{O}} \rangle_{N_s}$ is randomly drawn from a Gaussian distribution whose mean is the ideal expectation $\langle \hat{\mathcal{O}} \rangle$, while its maximum standard deviation is $\sigma_{max}(N_s) = 1/\sqrt{N_s}$.

The effect of this finite-size statistic is assessed by varying the order of magnitude of $N_s$ in the case of the STM task presented above. For each combination of delay value $\tau$ and number of samples $N_s$, we have optimized and tested the models in the same way as done in Sec. \ref{sec:memory}, keeping the same dataset sizes and exploring the hyperparameter space with the same criteria. From Fig. \ref{fig:Ns} we can clearly see that the prediction accuracy for both models increases with $N_s$ as a consequence of the better precision of the observable estimates. We have considered both a short ($\tau=2$) and a longer ($\tau=10$) delay.  For these two STM tasks, we observe that for a relatively small sample size, the FN model achieves higher capacities than the CD model, although the performance is not optimal for either. We also observe that the FN model can saturate its best performance for smaller samples (e.g. $N_s \sim 10^7$ for $\tau=10$) but in general, the CD model can outperform the FN model. Actually, even for a large delay, while the STM of the FN model saturates to rather poor values (for $N_s  \gtrsim  10^7$), the CD achieves a very good performance.

It is important to note that the relationship between the number of measurements and the performance found is specific to the model being considered, and can be assessed precisely only in some special cases \cite{garcia2022scalable}. While the precision of observable expectation values is expected to improve with a general trend of approximately $1/\sqrt N$, predicting how this translates into the performance in different tasks is more complex and cannot be done in general. As expected, the performance for averages obtained considering a sufficiently broad sample saturates to the ideal value, where averages are estimated over an infinite ensemble (represented by the rightmost values in Fig. \ref{fig:MG}).

In a real experimental implementation, due to the back-action effect of projective measurements after each sample is taken, the reservoir can no longer be used and a new experiment must be started, as analyzed in Ref. \cite{mujal2022time}. The fading memory property, which will be discussed in detail in section \ref{sec:universality}, can significantly reduce the experimental time required. In fact, it is not necessary to inject the whole time series in each experiment, but only a shorter number of points equal to the time width that the reservoir is able to remember. In addition, the experimental time can be further reduced by using a weak measurement protocol \cite{mujal2022time}. 
The relationship between continuous dissipation models and weak measurement schemes is an issue to be addressed in the future.

\subsection{Platforms}
Various driven-dissipative systems have been experimentally realized in numerous platforms such as cavity QED and cold atoms \cite{plat1,plat2,plat3}, circuit QED \cite{plat4}, and trapped ions \cite{plat5, plat6, Blatt2012}, due to the rapid progress of experimental techniques in the last 20 years. In addition, they are prominent platforms for quantum simulation and quantum computation \cite{Eng_diss}, especially relevant for quantum computing systems in the NISQ era.

In particular, several experimental platforms capable of implementing dissipative Ising models, which include our CD model, have been proposed in the literature. We mention significant results obtained with Rydberg atoms \cite{Rydb_ex1, Rydb_ex2, Rydb_ex3,Ryd_and_Ion}, ions traps \cite{Ryd_and_Ion, Ions_ex, Ions_ex2, Ions_ex3} and cold atoms \cite{cold_ex, cold_ex2}. For all these physical systems, the local independent dissipation present in Eq. \ref{Eq:ME_CD} can be easily implemented using standard techniques. Following the pioneering work of Ref. \cite{Eng_diss}, the desired dissipation can be achieved by coupling each qubit to an ancillary one that presents spontaneous emission through the well-known strategies of Ref. \cite{AtomsIons_transf}. Moreover, this model of dissipation can be easily implemented by means of optical pumping \cite{Optical_pumping}.

Beyond the analog system approach to QRC \cite{Markovic}, we also mention that quantum circuits can efficiently simulate generic Markovian dynamics, like that of Eq. \ref{Eq:ME_CD}. In particular, the strategy consists of implementing the collision model algorithm \cite{Cattaneo_collision} as it has been experimentally demonstrated on the IBM platform \cite{deco}.

\section{Discussion and Conclusions}\label{sec:conclusions}

In this paper, we have shown that tunable losses in external environments can be turned into a crucial factor for quantum reservoir computing, to tailor and optimize the memory capabilities of the system. In our analysis, we have compared the performances of continuous dissipation (CD) QRC maps with alternative ones based on a discontinuous erase-and-write, as proposed by Fuji and Nakajima (FN) \cite{Nakajima}. The CD reservoir is based on an input-dependent generator of the dynamics  $\bar{\mathcal{L}}(s_{k})$  and on the presence of Markovian dissipation modeled through a local master equation in the Lindblad form, with uniform and tunable decay rates.  

Through a set of non-linear memory and forecasting tasks commonly employed to benchmark time series processing,  we have shown that the degree of tunability of the losses represents a powerful tool that allows the system to be reconfigured with respect to the specific problem under consideration. Indeed, for each task, we have found a range of values of the losses for which the performance indicators exceed those obtained with the FN model, both for memory and temporal prediction tasks. This improvement was also shown to be robust in the realistic case of a finite number of measurement samples.

As a further major result, we have analytically shown that the CD model fulfills the three necessary conditions for time series processing as a reservoir computer (the echo state, fading memory and input separability properties) proving that it forms a class of universality, approximating any fading memory map with arbitrary precision. Finally, our proof has shown that this last result is a more general feature of Markovian dynamics in open quantum systems. In fact, universality is achieved if the generator of the dynamics has the following general, mild properties: for each input injection at sufficiently long $\Delta t$, it must admit only one stationary state; it must be a continuous function of the input; finally, for different inputs, the corresponding stationary states must be in turn different. 

Some considerations can be added about the generality and interpretation of our results. Looking at the FN and CD CPTP maps, we can say that the FN model of Eq. (\ref{eq:Naka}) can be described as a sequential map composed by a dissipation ($\Phi_D$) and a Hamiltonian evolution ($\Phi_U$): $\Phi = \Phi_U \circ \Phi_D$.  A slightly different approach was taken in Refs. \cite{QCirc2, NakaDiss} in the attempt to model the decoherent noise in a quantum circuit, where the roles of $\Phi_U$ and $\Phi_D$ were reversed: $\Phi = \Phi_D \circ \Phi_U$. The Markovian continuous-dissipation map of Eq. \eqref{eq:New_model} goes beyond these proposals as dissipation and driving act in a continuous and, in general, non-factorizable way. 

Interestingly, our map can approximate all these factorized ones \cite{Nakajima,QCirc2, NakaDiss} by a proper choice of the dissipation rates and their time dependence. Indeed, for the erase and write map of Eq. (\ref{eq:Naka}) this is shown in Appendix \ref{App:Approximation}. On the other hand, the model used in Refs. \cite{QCirc2, NakaDiss} assumes a noise modeled by a Markovian channel $\Phi_D$ so that also $\Phi$ is Markovian, and in general can be written in the GKLS form (\ref{eq:Linb}). In conclusion, we remark that the major generality and tunability obtained with our formalism has been revealed to be useful in practical applications as shown in the performance improvements found in Sec. \ref{sec:tasks}.

From a more physical point of view, we can relate the substantial improvements found in the CD model with the form of dissipation and driving (continuous information erasure and injection) in the reservoir. Indeed, an essential ingredient for RC to work is to find a balance between information spreading across the reservoir and dissipation, which is essential for the fading memory property. In the FN model, the input is injected locally and needs to spread during the evolution  \cite{martinez2021dynamical}, while dissipation is operated by the partial trace and not optimized. On the contrary, in the CD case, information is already injected over the entire reservoir, and at the same time, the degree of dissipation can be fine-tuned by the parameter $\gamma$. These two limiting features of the FN map are at the origin of the higher performance of the CD one, even varying the dimensions as shown in Appendix \ref{Sec:Scaling}.

While a clear improvement is already obtained here assuming the simplest form of dissipation, the idea of dissipation engineering can be explored also considering non-local losses \cite{cabot2018unveiling}, as well as non-Markovian dissipation \cite{nonMark}, opening the way to study a wider range of quantum reservoir computers. In conclusion, in this work, we have established a new paradigm of QRC based on Markovian dynamics that is suitable for generalization to more complex dissipation engineering techniques.

\section{Acknowledgments}
This work was partially supported by the María de Maeztu project CEX2021-001164-M funded by the  MICIU/AEI/10.13039/501100011033, by the QUARESC project (PID2019-109094GB-C21 and -C22/AEI/ 10.13039/501100011033) and by the COQUSY project PID2022-140506NB-C21 and -C22 funded by MCIU/AEI/10.13039/501100011033. This work has also been financially supported by the Ministry for Digital Transformation and of Civil Service of the Spanish Government through the QUANTUM ENIA project call - Quantum Spain project, and by the European Union through the Recovery, Transformation and Resilience Plan - NextGenerationEU within the framework of the Digital Spain 2026 Agenda. The CSIC Interdisciplinary Thematic Platform (PTI+) on Quantum Technologies in Spain (QTEP+) is also acknowledged. GLG is funded by the Spanish  Ministerio de Educaci\'on y Formaci\'on Profesional/Ministerio de Universidades and co-funded by the University of the Balearic Islands through the Beatriz Galindo program (BG20/00085). AS acknowledges funding from the CSIC hub on AI through the scholarship JAEIntroAIHUB2-19. The project that gave rise to these results received the support of a fellowship from the ”la Caixa” Foundation (ID 100010434). The fellowship code is LCF/BQ/DI23/11990081. RMP acknowledges the QCDI project funded by the Spanish Government and part of this work was also funded by MICINN/AEI/FEDER and the University of the Balearic Islands through a pre-doctoral fellowship (Grant No. MDM-2017-0711-18-1).

\newpage\appendix
\section{Fuji-Nakajima map and Lindblad dynamics}\label{App:Approximation}

\begin{figure}[t]
    \centering
    \includegraphics[width=\linewidth, keepaspectratio]{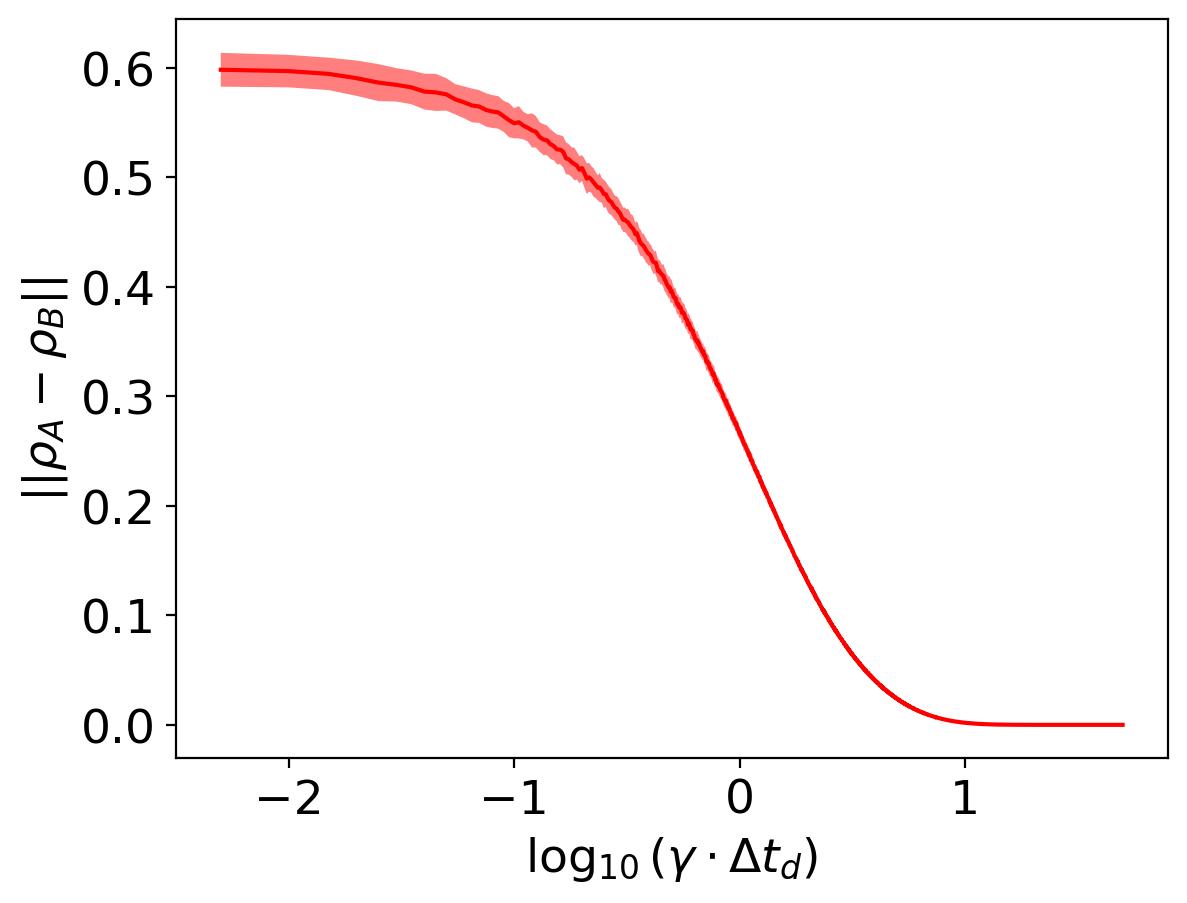}
    \caption{{Hilbert-Schmidt distance of 
    states evolving according to the left-hand side or the right-hand side of Eq. \eqref{eq:Approx}, 
    say $\rho_{A}$ and $\rho_{B}$, as a function of the amount of  dissipation. Values are averaged over 100 random pairs of density matrices and inputs, respectively $\rho_k$ and $s_k$ in the formula. }}
    \label{fig:Dist}
\end{figure}

In the following, we want to frame the FN model into our Markovian strategy to obtain dissipation by designing an equivalent input injection strategy. The idea is to replace the role played by the partial trace by introducing a strong loss on the first qubit and then preparing its state by applying the rotation operator 
\begin{align*}
R_y(\theta_k)= \begin{pmatrix}
\cos\theta_k & \sin\theta_k \\
-\sin\theta_k & \cos\theta_k 
\end{pmatrix}
\end{align*}
to it. More precisely the network's damping rates will be
\begin{equation*}
    \gamma_{i}=\begin{cases} \gamma , & \mbox{if $i=1$ } \\ 0, & \mbox{otherwise}
    \end{cases}
\end{equation*}
and the input $s_{k}$ will be converted into an angle of rotation $\theta_{k}$ according to $\theta_{k}=\arccos(\sqrt{1-s_k})$. Accounting for the two steps, we can determine the conditions for the entire process, at each input injection, to be equivalent to the updating step of Eq. \eqref{eq:Naka}, i.e.,
\begin{equation} \label{eq:Approx}
  R_{y}^{(1)}(\theta_{k+1})
  e^{\mathcal{D}_{1}\Delta t_d} \rho_{k} \approx \rho_{k+1}^{(1)}\otimes\Tr_{(1)}\{\rho_{k}\},
 \end{equation}

\section{Proof of universality}\label{sec:universality}

 In this section, we will prove that the CD model proposed in Eq. \eqref{eq:New_model} forms a universality class for reservoir computing. In the context of QRC, universality corresponds to an approximation property and, in particular, is referred to the capability to arbitrarily well approximate any fading memory map \cite{LSM,grigoryeva2018echo,Qcircuit,nokkala2021gaussian}. It will be shown that our proposal satisfies both the echo state property (ESP) and the fading memory property (FMP). Finally, we will prove separability verifying that the hypotheses of the Stone-Weierstrass theorem are fulfilled, then proving the universality. 
 
\subsection{Echo state property}\label{Subsec:ESP}
 The ESP refers to the capability of a reservoir to forget the initial conditions in the limit of an infinite input sequence \cite{EC_prop}. It has been recently shown that this is equivalent to the usual definition of a unique solution in the RC literature, under some mild conditions (Theorem 1 in \cite{manjunath2022embedding}). In our case of study, it means that given a left infinite and time-ordered input sequence $\{\cdots, s_{-1},s_{0} \}$ (where the 0-th time step corresponds to the last input) and given two different initial quantum states $\rho_1$ and, $\rho_2$, whose dynamics is driven by this sequence, the evolved states will become indistinguishable:
\begin{equation}
 \label{Eq:Echo}
    \lim_{k \to -\infty} \| \prod_{i=k}^{0} e^{\bar{\mathcal{L}}(s_{k})\Delta t} \left( \rho_1 - \rho_2 \right) \| = 0, 
\end{equation}
where the limit is considered to be pointwise and $\| \cdot \| $ indicates any matrix norm (since all norms are equivalent in finite-dimensional spaces). 

A sufficient condition for this relation to hold is strict contractivity (Theorem 2.2 in Ref. \cite{chen2022nonlinear}):
$||e^{\bar{\mathcal{L}}(s_{k})\Delta t}(\rho_1-\rho_2)||\leq r||\rho_1-\rho_2||$
for all $s_k$ and any pair of states $\rho_1$ and $\rho_2$, where $0\leq r<1$. 
In particular, this can be ensured for Markovian maps that provide a unique stationary state for a generic input $s_k$, a property that our model possesses according to Theorem 2 \cite{Unique}, setting a long enough value of $\Delta t$. More precisely, for a generic generator $\bar{\mathcal{L}}(s_{k})$, we can always define a finite mixing time, denoted as $\Delta t_{mix}$, such that if $\Delta t \ge \Delta t_{mix}$ then the map is strictly contractive and the ESP holds. 

As a further observation, we estimated the $\Delta t_{mix}$ trend as the system dimensions increase. Our numerical analysis suggests that the $\Delta t_{mix}$ upper bound has an asymptotic behavior $\sim \frac{N}{\gamma}$ (see Appendix \ref{App:MixTime} for more details).
Concluding, the mixing time of all the proposed maps $e^{\bar{\mathcal{L}}(s_{k})\Delta t}$ in addition to being finite, according to what we have observed numerically, scales efficiently with the number of qubits.

\subsection{Fading memory}
Considering the set of left infinite sequences of inputs belonging to the interval $[0,1]$,  $K^{-}([0,1])$, the fading memory property is a condition of continuity of functionals defined on this set with a given norm. For a sequence $s \in K^{-}([0,1])$, this norm is defined starting from a generic null sequence, say $w=\{w_k\}_{k\geq 0}$ such that $\lim_{k\rightarrow\infty}w_k=0$, in the following way:
\begin{equation*}
    \| s \|_{\textit{w}}=\sup_{k \in \mathbb{Z}^{-}}|s_{k}|w_{-k},
\end{equation*}
where $s_{k}$ and $w_{-k}$ are respectively elements of s and \textit{w} and $\mathbb{Z}^{-}$ is the set of negative integers.
By definition, we say that a map has fading memory if, for any null sequence \textit{w}, it is continuous in $(K^{-}([0,1]),\| \cdot \|_{\textit{w}})$. In order to prove that the model of Eq. \eqref{eq:New_model} has fading memory, it is sufficient to prove that $e^{\bar{\mathcal{L}}(s_{k})\Delta t}$, with $\Delta t > \Delta t_{mix}$ to ensure the ESP, is a continuous function of $s_{k}$ according to \cite{Chen_2019} (Lemma 3). If $\bar{\mathcal{L}}(s_{k})$ is continuous then the continuity of its exponential will be a direct consequence. Let $s_{k}$,$u_{k}$ $\in [0,1]$, $\| \cdot \|_{2}$ be the Hilbert-Schmidt norm and  $\rho$ be a unit density matrix belonging to $\mathbb{C}^{ 2^{N} \times  2^{N} } $. Then
    \begin{align}
    \sup_{\rho} \left \| (\bar{\mathcal{L}}(s_{k}) - \bar{\mathcal{L}}(u_{k})) \rho  \right\|_{2} =& \nonumber\\     \sup_{\rho} \left\|(s_{k}-u_{k}) \left[\sum_{i=1}^{N}\sigma_{i}^{x},\rho\right]     \right\|_{2}  \le 2 \cdot N \cdot |s_{k}-u_{k}| \label{eq:FM} 
    \end{align}
where the first inequality is a consequence of the following property of the commutator: $\left\|\left[\sigma_{i}^{x},\rho\right]\right \|_{2} \le 2 \cdot \| \rho \|_{2} = 2$ which can be proven expanding $\rho$ as a linear combination of Pauli strings and using their commutation rules. The last term of Eq. (\ref{eq:FM}) implies continuity because it shows that an arbitrarily small distance between the two functions, let us say $\epsilon$, is reached when the distance of the two inputs is $|s_{k}-u_{k}| < \delta_{\epsilon}=\frac{\epsilon}{2 N}$.

 We then conclude that the family, say $\mathfrak{L}$, of functionals defined by Eq. \eqref{eq:New_model}, with the output layer of expected values,  belongs to the set of fading memory functionals $C(K^{-}([0,1]),\| \cdot \|_{\textit{w}})$, which map input sequences into real expected values of observables.
 
\subsection{Stone-Weierstrass theorem}
The universality condition is obtained if $\mathfrak{L}$ is also a dense subset of $C(K^{-}([0,1]),\| \cdot \|_{\textit{w}})$. It implies that any fading memory map $f \in C(K^{-}([0,1]))$ can be arbitrarily well approximated by an element of $\mathfrak{L}$. Quantitatively for any $f \in C(K^{-}([0,1]))$ and for any $\epsilon > 0$ exist $l \in \mathfrak{L}$ such that the following condition holds: 
\begin{equation*}
\left|f-l\right|_{\infty} = \sup_{s \in K^{-}([0,1])} \left|f(s)-l(s)\right|< \epsilon.  
\end{equation*}
A sufficient condition to obtain this result is given by the well-known \textbf{Stone-Weierstrass theorem}: \textit{Let E be a compact metric space and $\mathbf{C}$(E) be the set of real-valued continuous
functions defined on E. If a subalgebra A of $\mathbf{C}$(E) contains the constant functions and separates points of E, then A
is dense in $\mathbf{C}$(E)}.

A first important observation is that the input space $(K^{-}([0,1]),\| \cdot \|_{\textit{w}})$ is a compact space according to \cite{Ortega} (Lemma 2) as required. 

\subsection{Input separability}
Now we prove the separability of $\mathfrak{L}$. It means that given two sequences $s, \bar{s} \in K^{-}([0,1])$ such that $s \ne \bar{s}$ exists at least one element of $\mathfrak{L}$ able to separate them. In our context it is translated to the condition that the corresponding density matrices at the 0-th time step must be different:  $\left \| \rho_0 - \bar{\rho}_0 \right \| \ne 0$.

It is useful at this point to show an important property of Eq. \eqref{eq:New_model}:

\textbf{Lemma 1}: \textit{Given two different inputs $s_k$, $u_k$ then 
the corresponding stationary states of $\bar{\mathcal{L}}(s_k)$ and $\bar{\mathcal{L}}(u_k)$ are different}.

\textit{Proof:}

We firstly give a necessary condition for a stationary state $\rho_{ss}$ which corresponds to a generator $\bar{\mathcal{L}}(s_k)$. It is helpful to expand $\rho_{ss}$ in the basis of Pauli strings:
\begin{equation*}
    \rho_{ss}=\sum_{i_{1}, \cdots, i_{N}} \Tr{\sigma^{i_1} \otimes \cdots \otimes \sigma^{i_N} \cdot \rho_{ss}} \sigma^{i_1} \otimes \cdots \otimes \sigma^{i_N}
\end{equation*}
where the indexes $\{i_{n}\}$ label single Pauli matrices: $i_{n}=x,y,z$ or the identity matrix, say $\sigma^{0}=\mathbb{I}$, for the Hilbert space of a single qubit.
A necessary condition can be found projecting the definition of $\rho_{ss}$ (i.e. $\bar{\mathcal{L}}(s_k)\rho_{ss}=0$) on a given Pauli string; we will consider $\sigma^{z}_{i}$:
\begin{equation}
    \label{Eq:stat_nec_1}
    \Tr{\sigma_{i}^{z}\bar{\mathcal{L}}(s_k)\rho_{ss}}=0.
\end{equation}
Observing that we can write the Lindbladian of Eq. $\eqref{eq:New_model}$ as a sum of single qubit dissipators:
\begin{equation*}
    \mathcal{D} = \gamma \sum_{i=1}^{N} d_i
\end{equation*}
where $d_{i}(\rho)= \sigma_{i}^{-} \rho \sigma_{i}^{+}-\frac{1}{2} \{\sigma_{i}^{+}\sigma_{i}^{-},\rho \}$ with $\sigma^{+}=\sigma^{-\dagger}$, its action on the single qubit Hilbert space:
\begin{equation*}
  \begin{aligned}
  d_{i}(\mathbb{I})&=-\sigma^{z}\\
    d_{i}(\sigma^z)&=-\sigma^{z}\\
  d_{i}(\sigma^y)&=-\frac{1}{2}\sigma^{y}\\
  d_{i}(\sigma^x)&=-\frac{1}{2}\sigma^{x}\\
  \end{aligned}  
\end{equation*}
implies the following expression for Eq. \eqref{Eq:stat_nec_1}:
\begin{equation}
    \label{Eq:stat_nec_2}
    \gamma \cdot \alpha_i^z+\frac{\gamma}{2^{N}} - \sum_{j\ne i} J_{i,j}\cdot 2 \cdot \alpha_{i,j}^{y,x} - 2 \cdot h \cdot(s_k+1) \cdot \alpha_{i}^y=0,
\end{equation}
where $\alpha_i^z=\Tr{\sigma_i^z \rho_{ss}}$, $\alpha_{i,j}^{y,x}=\Tr{\sigma_{i}^y\sigma_j^x \rho_{ss}}$, $\alpha_i^y=\Tr{\sigma_i^y \rho_{ss}}$.

Considering now another generic input $u_k$, such that $u_k \ne s_k$, if $\rho_{ss}$ is a common stationary state of both, the following relation must be satisfied: 
\begin{equation*}
    \bigl( \bar{\mathcal{L}}(s_{k})-\bar{\mathcal{L}}(u_{k}) \bigr)\rho_{ss}=0,
\end{equation*}
implying that 
\begin{equation}
    \label{Eq:comm}
    \left[ \sum_{i=1}^{N}\sigma_{i}^{x},\rho_{ss}\right] = 0. 
\end{equation}
Equation \eqref{Eq:comm} gives necessary conditions for the coefficients of $\rho_{ss}$ and, among these, it gives these relations: $\Tr{\sigma_i^a \rho_{ss}}=\alpha_i^a=0$, $\Tr{\sigma_{i}^{a}\sigma_{j}^{x} \rho_{ss}}=\alpha_{ij}^{ax}=0$ with $a=y,z$ and $1 \le i,j \le N$.  
Then Eq. \eqref{Eq:stat_nec_2} becomes $\frac{\gamma}{2^{N}}=0$ which is not satisfied because we are assuming that $\gamma \ne 0$ in order to guarantee the ESP. As a consequence, we can conclude that Lemma 1 holds. $\blacksquare$

Another useful statement to arrive at the input separability is the following:

\textbf{Lemma 2}: \textit{Let $s_k$ and $\rho_{ss}(s_k)$ be a generic input and its corresponding unique stationary state, let $\mathbb{P}$ be the set of the density matrices of the considered N qubits system and let $d\colon \mathbb{P} \times \mathbb{P} \to \mathbb{R}^+$ be the distance induced by the Hilbert-Schmidt product, then the following function $g\colon \mathbb{R}^+\times \mathbb{P} \to \mathbb{R}^+,\quad(t,\rho) \mapsto d\left(e^{\bar{\mathcal{L}}(s_{k})t}\rho,\rho_{ss}(s_k)\right)$ is bounded, its maximum $\left(d_{max}^{(s_k)}\right)$ always exists and it is strictly decreasing in t when $t>\Delta t_{mix}$}.

\textit{Proof:}

We first observe that for any fixed value of the time g is a continuous function on $\mathbb{P}$. This is because it is a composition of continuous functions: 
the action of $e^{\bar{\mathcal{L}}(s_{k})t}$ composed to the action of the distance from a fixed point. As a consequence because $\mathbb{P}$ is a compact set and because the continuous image of a compact set is a compact set we arrive at the following relation:
\begin{equation}
    \label{Eq:g_bound}
    \left.g(\mathbb{P})\right|_{t} = \left[0, d_{max}^{(s_k)}(t)\right].
\end{equation}
We observe that for the boundaries of the set of the density matrices $d_{max}^{(s_k)}(t) \le 2$. Considering now the action of $g$ for a fixed density matrix in $\mathbb{P}$, we know that $\left.g(t)\right|_{\rho} = d(e^{\bar{\mathcal{L}}(s_{k})t}\rho, \rho_{ss}(s_k))$ is strictly decreasing for all $\rho \ne \rho_{ss}(s_k)$ if $t>\Delta t_{mix}$ as a consequence of the already mentioned strict contractivity of $e^{\bar{\mathcal{L}}(s_{k})t}$. Finally the same property must hold for $d_{max}^{(s_k)}(t)$ while $\lim_{t \to +\infty} d_{max}^{(s_k)}(t)=0$ and $d_{max}^{(s_k)}(0)=2$. $\blacksquare$

We can now return to the problem of the separability in which we have the two generic different sequences $s$ and $\bar{s}$. Considering the smallest index J such as $s_{-J} \ne \bar{s}_{-J}$, applying Lemma 1 we know that two different stationary states exist: $\rho_{ss}(s_{-J})$ and $\rho_{ss}(\bar{s}_{-J})$. We define two open balls of radius \textit{r} respectively centered on the two states: $B_{r}(\rho_{ss}(s_{-J}))$ and $B_{r}(\rho_{ss}(\bar{s}_{-J}))$ such that 
$B_{r}(\rho_{ss}(s_{-J})) \cap B_{r}(\rho_{ss}(\bar{s}_{-J})) = \emptyset$. The last condition is easily satisfied if $r < d(\rho_{ss}(s_{-J}), \rho_{ss}(\bar{s}_{-J}))/2$. In order to find an element of $\mathfrak{L}$ which separates the sequences, applying Lemma 2, we have to set $\Delta t$ into a sufficiently long value so that: $d_{max}^{(s_{-J})}(\Delta t),d_{max}^{(\bar{s}_{-J})}(\Delta t) < r$.

With this condition, it is ensured that after the applications of the two inputs, the resulting states of the reservoirs will be different regardless of their states at the time-step $-J$. After this injection by hypothesis, the subsequent inputs will be the same, and necessary the corresponding states of the reservoir at the time 0 will be different because $e^{\bar{\mathcal{L}}(s_{k})t}$ is a full rank linear operator. We can in this way conclude that the input separability is satisfied.
\subsection{Polynomial algebra and final considerations about the universality}
Another hypothesis that must be satisfied in order to have universality is the presence of a subalgebra in $\mathfrak{L}$. Since we are working on a system that gives as output a linear combination of observables of the density matrix we have to look for a polynomial algebra.
 It can be obtained by adding to the family the model of Eq. $\eqref{eq:New_model}$ in spatial multiplexing. It means that we can consider V different and independent states $\{\rho^{(1)},\cdots , \rho^{(V)} \}$ whose dynamics will be lead by generators of the form of Eq. $\eqref{eq:New_model}$ with, in general, different value of the characteristic hyperparameters: $\{\bar{\mathcal{L}}_1, \cdots, \bar{\mathcal{L}}_V\}$. We can write the total updating rule of the total reservoir in the following way:
\begin{align*}
    \rho_{k+1}^{tot} &= \rho^{(1)}_{k+1} \otimes \cdots \otimes \rho^{(V)}_{k+1} \\
    &= e^{\bar{\mathcal{L}}_1(s_{k+1}) \Delta t} \cdots e^{\bar{\mathcal{L}}_V(s_{k+1}) \Delta t} \rho^{(1)}_{k} \otimes \cdots \otimes \rho^{(V)}_{k}.
\end{align*}
Considering the set of the polynomial outputs for all the single states $\{ \mathcal{P}_{1}, \cdots , \mathcal{P}_{V}\}$ we achieve the algebra considering as output for the reservoir a linear combination of the polynomials:
\begin{equation*}
    \mathcal{P}_{tot}=\beta_0 + \sum_{i=1}^{V} \beta_i \cdot  \mathcal{P}_{i}.
\end{equation*} 
These newly added reservoirs satisfy the fading memory condition according to \cite{Chen_2019} (Lemma 5).

The only condition that remains to be proven in order to assert the Stone-Weierstrass theorem is the presence of constant functions in the family $\mathfrak{L}$ but it is trivially satisfied due to the fact that we are working with polynomial inputs.
We can now conclude that $\mathfrak{L}$ is a universal class for reservoir computing.

Finally, we notice that this proof does not hold only for Eq. \eqref{eq:New_model} but 
for a more general class of reservoirs working with a master equation. 
The conditions required for the generator $\mathcal{L}$ are the following: 
(i) it admits only one stationary state for each input $s_k$, (ii) it must be a continuous function of $s_k$, (iii) given two different generic inputs, say $s_k$ and $u_k$, the stationary states of  $\bar{\mathcal{L}}(s_{k})$ and  $\bar{\mathcal{L}}(u_{k})$ are also different. 

\section{Mixing time scaling }\label{App:MixTime}
\begin{figure}[t]
    \centering
    \includegraphics[width=\linewidth, keepaspectratio]{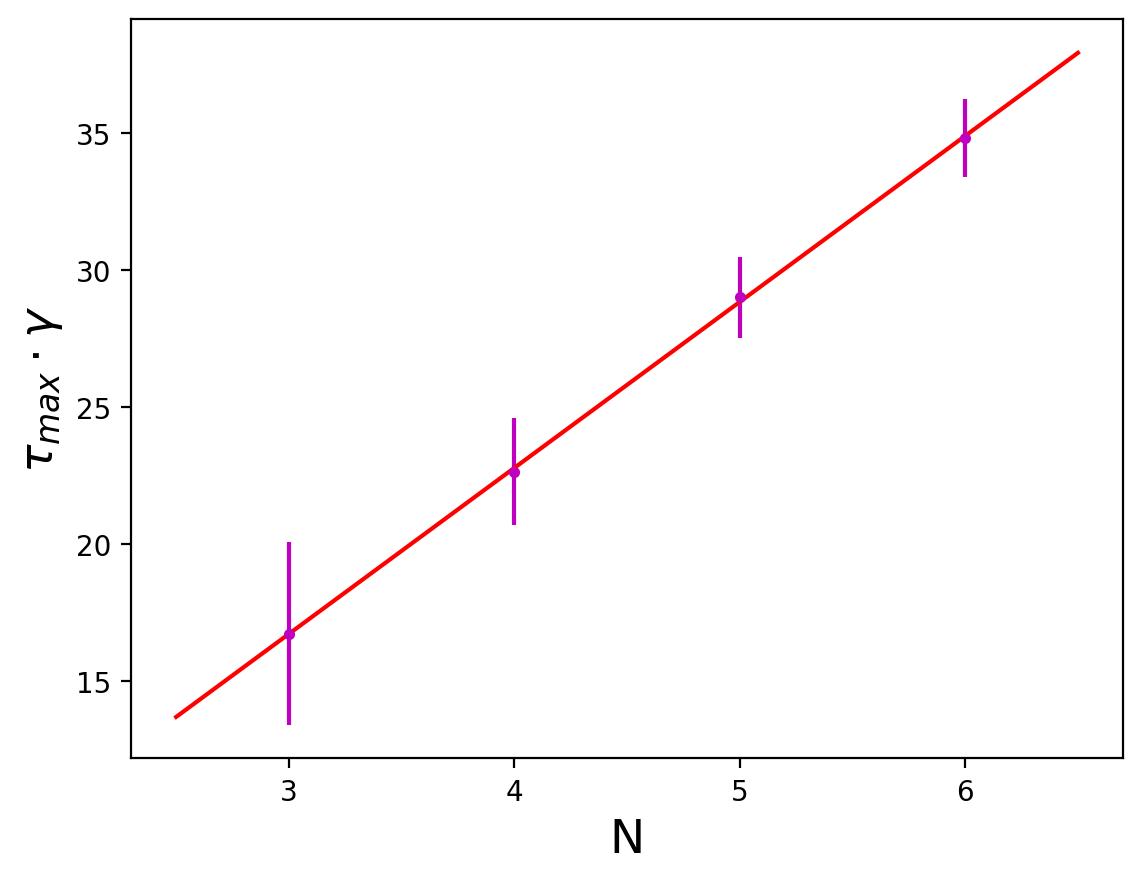}
    \caption{Max average values of $\tau$ in $\gamma$ units as a function of the number of spins. The purple points indicate the averages while the purple vertical lines cover one standard deviation taken as statistical error over the $\{J_{i,j}\}$ realizations. The linear fit result is represented by the red line.}
    \label{fig:Mix_t}
\end{figure}
In Sec. \ref{Subsec:ESP}, we have defined the mixing time $\Delta t_{mix}$ as the minimum interval of time $\Delta t$ required of $e^{\bar{\mathcal{L}}(s_{k})\Delta t}$ to fulfill the Echo State Property. From a computational point of view, we are interested in estimating how $\Delta t_{mix}$ scales with the system size. We will now show the strategy used to achieve it through numerical simulations. We, firstly, recall that $e^{\bar{\mathcal{L}}(s_{k})\Delta t}$ can always be expanded in terms of its dual basis except for a countable number of points in the hyperparameter space that have zero probability to occur \cite{Lindb_diag}. Let $\{\lambda_i\}$ be the set of eigenvalues of $\bar{\mathcal{L}}(s_{k})$ and let $\{| r_i \rangle \! \rangle\}$ and $\{| l_i \rangle \! \rangle\}$ be the corresponding set of orthogonal and normalized right and left eigenvectors respect the Hilbert-Schmidt product. As a consequence of the uniqueness of the stationary state of $\bar{\mathcal{L}}(s_{k})$ (Theorem 2 of Ref. \cite{Unique}), when its action is restricted to the set of traceless Hermitian matrices, which we will denote as $\mathcal{H}_0$, we can conclude that the real part of all its eigenvalues will be strictly less than zero. For the sake of definition, we will order the eigenvalues such that $\Re{\lambda_0}=0 > \Re{\lambda_1} \ge \dots \ge \Re{\lambda_{4^N-1}}$ arriving at the identity:
\begin{align*}
e^{\bar{\mathcal{L}}(s_{k})\Delta t}|_{\mathcal{H}_0} &= \sum^{4^N-1}_{i=1} e^{\lambda_i \Delta t} \frac{| r_i \rangle \! \rangle   \langle \! \langle l_i |}{ \langle \! \langle l_i | r_i \rangle \! \rangle } \\
&= \sum^{4^N-1}_{i = 1} c_i \cdot e^{\lambda_i \Delta t} | r_i \rangle \! \rangle \langle \! \langle l_i |
\end{align*}
where N is the number of spins of the model and $c_i = 1/\langle \! \langle l_i | r_i \rangle \! \rangle$. The ESP is ensured when the operator norm of $e^{\bar{\mathcal{L}}(s_{k})\Delta t}|_{\mathcal{H}_0}$ is strictly less then one. Considering that it is induced by the Hilbert-Schmidt norm, we find:
\begin{align}
    \left\|e^{\bar{\mathcal{L}}(s_{k})\Delta t}|_{\mathcal{H}_0}\right \|_{2-2} &= \left \|\sum^{4^N-1}_{i = 1} c_i e^{\lambda_i \Delta t} | r_i \rangle \! \rangle   \langle \! \langle l_i | \right \|_{2-2} \nonumber \\
     &\le \sum^{4^N-1}_{i = 1} |c_i| \cdot |e^{\lambda_i \Delta t}| \nonumber \\
     &\le 4^N \cdot |c_M| \cdot e^{\Re{\lambda_1} \Delta t} \label{eq:ESP_mix}
\end{align}
where $|c_M| = \max_i |c_i|$. From Eq. (\ref{eq:ESP_mix}), we can find the order of magnitude for the upper bound of the mixing time, calling it as $\tau$, by definition the following relation holds: $\Delta t_{mix} \apprle \tau$. Writing $|c_M|=e^{\eta}$, we arrive to the expression: $\tau = \frac{N}{|\Re{\lambda_1}|} + \frac{\eta}{|\Re{\lambda_1}|}$.
We have numerically computed a maximal value of $\tau$ spanning the system dimension from 3 to 6 spins. For each case, the magnetic field $h$ took the values from the set \{0.01, 0.05, 0.1, 0.5, 1,5, 10\}, the damping rate $\gamma$ from \{0.01, 0.1, 1, 10\} while the input has been fixed to be $s_k = 0$. For all the possible combinations we have simulated 100 realizations of the Hamiltonian couplings $\{J_{i,j}\}$ and we have taken the average value of $\tau$ as a representative. In our analysis, we have selected the max average for each N, which computationally corresponds to the worst case, in order to estimate an upper bound. As shown in Fig. \ref{fig:Mix_t}, our numerics suggest that the max order of magnitude of $\Delta t_{mix}$ scales proportionally to $\frac{N}{\gamma}$.  
 As a result, we have numerically observed that the minimum value of $\Delta t$ required to reach the ESP, for the map of our model generated by $\bar{\mathcal{L}}(s_{k})$, scales in an efficient way with respect to the number of qubits.

\section{Training details}\label{Sec:training}

As explained in Sec. \ref{sec:experiments}, 
we benchmarked the considered quantum reservoirs by computing their performance for different network realizations. In the case of memory tasks, we also varied the random inputs injected. It is important to note that training a reservoir for a specific task depends on the reservoir evolution rule. Therefore, we performed independent training for each different network coupling sampled.

Going more into details, we indicate with $\{ \langle \hat{\mathcal{O}}\rangle _{k,j} \}_{k=1,j=1}^{L,M}$ the set of the expectation values of the reservoir observables involved in the training phase, where $L$ is the number of training points and $M$ the number of chosen observables. 

Considering the free weights that determine the reservoir output at each time $\{w_j\}_{j=1}^{M}$, to be determined by the training, and set of targets which defines the task of interest $\{y_k\}_{k=1}^{L}$, the implemented training consisted on minimizing the following euclidean distance:
\begin{align*}
\sqrt{\sum_{k=1}^{L}\Big(y_k - \sum_{j=1}^{M} w_j \langle \hat{\mathcal{O}}\rangle _{k,j}\Big)^2}.
\end{align*}
We have numerically performed this optimization by using the LAPACK library \cite{Anderson1999}, whose strategy consists of making use of the singular value decomposition.

\section{Performances for different reservoir sizes}\label{Sec:Scaling}
\begin{figure}[t]
    \centering
    \includegraphics[width=\linewidth, keepaspectratio]{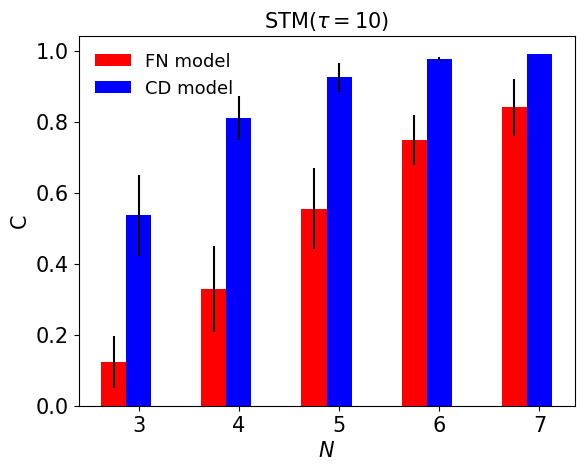}
    \caption{{STM task optimal performances of the FN model and the CD one as a function of the number of qubits $N$, fixing a delay $\tau = 10$. The optimal values have been computed in the same way as in Sec. \ref{sec:tasks}.  We averaged over 100 network realizations and input series for the cases of $N \leq 5$ while we took a statistics of 10 realizations for the complementary ones. The error bars refer to one standard deviation taken as a statistical error.}}
    \label{fig:STM_scal}
\end{figure}

While in the main text we considered five qubit reservoirs, here we explore the effect of increasing or decreasing the size of the reservoir. Following the same optimization procedure described in Sec. \ref{sec:tasks} regarding the choice of optimal hyperparameters, in Fig. \ref{fig:STM_scal} we show the optimal STM task performances of the FN model and the CD model and for a delay $\tau = 10$.  

The dimension of the output layer changes with the system size. Thus, as expected, the performance of both the FN and CD models improves with the number of qubits (more degrees of freedom are exploited for solving the task). Interestingly, the CD model outperforms the FN model for all considered values of $N$. For the task at hand, it is reasonable to expect that for a sufficiently large reservoir, both QRCs are able to achieve optimal performance. However, we see that the CD model 
requires a smaller number of qubits to saturate. In fact, $N=6$ qubits are already sufficient to reach a capacity value $C \simeq 1$, while the FN would require $N > 7$, implying the need for more resources to achieve the same performance.

\printbibliography

\end{document}